\documentclass[12pt]{article}
\usepackage{amsmath,amssymb,color}

\newcommand{\bea}{\begin{eqnarray}}
\newcommand{\eea}{\end{eqnarray}}
\newcommand{\be}{\begin{equation}}
\newcommand{\ee}{\end{equation}}
\newcommand{\vs}[1]{\vspace{#1 mm}}

\newcommand{\dsl}{\pa \kern-0.5em /}

\newcommand{\pa}{\partial}

\newcommand{\nn}{\nonumber\\}

\begin{document}
\topmargin 0mm
\oddsidemargin 0mm

\begin{flushright}

USTC-ICTS/PCFT-24-14\\

\end{flushright}

\vspace{2mm}

\begin{center}

{\Large \bf The open string pair production revisited}

\vs{10}

{\large  J. X. Lu}

\vspace{4mm}

{\em  Interdisciplinary Center for Theoretical Study\\
 University of Science and Technology of China, Hefei, Anhui
 230026, China\\
 \medskip
 Peng Huanwu Center for Fundamental Theory, Hefei, Anhui 230026, China\\

}

\end{center}

\vs{10}

\begin{abstract}
The experimental efforts in testing the QED vacuum properties such as the Schwinger pair production in the presence of a strong electric field have so far not been successful. This raises a potential possibility regarding if the usual QED vacuum picture is a complete one. In this paper, we address this possibility by taking our own (1 + 3)-dimensional world as a visible D3 brane with a nearby (hidden) D3, placed parallel at a separation, in Type IIB superstring theory and by considering an analogous open string pair production process. This setup can be taken as a simplified version of the underlying QED resulting from the particle physics standard model constructed from the intersecting D-branes. We will use this simple setup to demonstrate that the stringy pair production rate in the weak-field limit, though as expected to equal to the corresponding QED rate from the braneworld picture, contains far more important information than the QED one and as such the future detection of the Schwinger pair production can teach us lessons about the existence of extra dimension(s) and a potential source of dark matter among other things.
 \\
\end{abstract}

\newpage

\section{Introduction}
The great success of quantum electrodynamics (QED) appears to make nobody doubt its underlying vacuum picture.  There are ways to test such a picture. For example, by applying a constant strong electric field to the vacuum, we expect the Schwinger pair production to occur \cite{Schwinger:1951nm}.  It is generally  believed that if the applied constant electric field reaches its threshold value which can be estimated as $e E_{T} \sim m^{2}_{e} \to E_{T} \sim 10^{18}$ Volt/m with $m_{e}$ the mass of electron or positron, one would detect such pair production. The null direct experimental detection of this pair production so far is largely attributed to the smallness of the electric field which can be realized in laboratory\footnote{The largest constant electric field which can be realized in laboratory is believed to be on the order of $10^{10}$ Volt/m. We learned this from our experimental colleague Zhengguo Zhao. The strongest direct-current magnetic field generated is on the order of 50 Tesla, see \cite{smf}, for example.}.  Even in the case in which there is a large enough electric field such as in quark gluon plasma (QGP)  experiment, there is no direct experimental evidence for this pair production though it is believed to occur and the lack of its detection is attributed to the large background  which overtakes this pair production. 

Attaining this threshold electric field in laboratory relies on ever evolving laser technologies. Though feasible eventually, it appears hardly experimentally accessible in the not too distant future.  An alternative approach is to use heavy nuclei\footnote{The author would like to thank Xu-Guang Huang for bringing his attention to this.}  as a source of strong electric field based on the pioneering work \cite{PS1945, GZ1969, PG1969}.  The basic idea is that when the nuclear charge $Z > Z_{0} = 137$,  the energy $E (Z)$ for the 1s state of hydrogen-like ions from the relativistic quantum mechanical Dirac equation decreases to the negative region and eventually crosses  the value  $- m_{e} c^{2}$ at the critical $Z_{\rm cr} \approx 173$ \cite{PS1945, GZ1969, PG1969, ZP1972, GMR1985, MS1994, GMV2017}. After this crossing, this level ``dives`` into the negative-energy Dirac continuum and becomes a resonance. If this supercritical resonance state is initially vacant, it can then  be occupied  by two electrons  from the negative-energy continuum  with the emission of two positrons\cite{GZ1969, PG1969, ZP1972, GMR1985, MS1994}.  The superheavy nuclei with $Z \ge Z_{\rm cr}$ can be formed via two heavy colliding ions.  The observation of the electron-positron pairs spontaneously produced during the collision would be the  direct evidence of  vacuum decay, therefore a test of QED vacuum picture. However, no sign of this has been found \cite{MS1994, ahmad1999}. There are proposals for investigating the  supercritical collisions at the upcoming accelerator facilities \cite{exp1-2009, exp2-2015, exp-3-2017}, allowing to perform experiments on an entirely new level.  New theoretical efforts for this purpose have also been put forward recently (see, for examples, \cite{Maltsev:2019ytv, Popov:2020xmd} and references therein).

Given what has been said, however, the null direct detection of the electron-positron pairs  so far could also imply the other possibility regarding if the usual QED vacuum picture is a complete one since after all from the modern view, QED is just an effective description of the low energy physics of the underlying UV complete fundamental theory.  We will explore this possibility using the open string pair production in Type II superstrings studied recently by the present author and his collaborators  in this paper.

For this, we consider our (1 + 3)-dimensional world as a D3 brane.  Unlike in QED, an isolated Dp brane in oriented Type II superstring theories, even it carries a constant electric flux, cannot in general give rise to the open string pair production as discussed in \cite{Lu:2018nsc, Jia:2018mlr, Lu:2020hml} and further elaborated recently in  \cite{Lu:2023jxe, Lu:2023sag} by the present author and  his collaborators. The reason is simple that the applied electric field can only stretch the two ends, carrying equal but opposite charges, of each of the virtual open strings unless the field reaches its critical value, breaks the open string and makes the Dp brane itself unstable\footnote{However, this is not the case for un-oriented  bosonic string or un-oriented Type I superstring and the corresponding open string pair production was discussed a while ago by Bachas and Porrati in \cite{Bachas:1992bh, Porrati:1993qd}. In the holographic case, the Schwinger effect had been discussed firstly in \cite{Gorsky:2001up, Semenoff:2011ng} for which the strong coupling can be studied using AdS/CFT and in the second reference discussed was an interesting phenomenon that the suppression exponential in the pair production rate disappears when the applied electric field reaches beyond the critical one.}. This critical electric field is hardly possible in practice and further we don't want the brane or the world we live itself unstable.  

The simplest system that can give rise to the open string pair production,  in the spirit of Schwinger pair production in QED \cite{Schwinger:1951nm}, consists of two Dp branes in Type II superstring theories as pursued in \cite{Lu:2018nsc, Jia:2018mlr, Lu:2017tnm, Lu:2018suj}.    Here the two Dp-branes  are placed parallel at a separation with each carrying a constant worldvolume electric flux\footnote{\label{fn3} Adding a worldvolume magnetic field on either of the D3 branes  \cite{Lu:2017tnm,  Lu:2018nsc, Jia:2018mlr, Lu:2019ynq, Lu:2020hml} can enhance the rate and  in certain cases  \cite{Lu:2019ynq, Lu:2020hml},  a sizable one can be achieved.}.  In this paper, we specify to $p = 3$.

Though this system is not a realistic one,  it captures the essential feature\footnote{The author would like to thank Tianjun Li for discussion and for explaining the spontaneous symmetry breaking from the perspective of the intersecting branes.} of the underlying QED vacuum picture obtained from the particle physics standard model constructed from the intersecting D6 branes (for examples, see \cite{Ibanez:2001nd, Cvetic:2001nr, Blumenhagen:2005mu}).  Note that the intersecting D6 branes in this construction have a common large (1 + 3)-dimensional world while the other two spatial dimensions are compact and intersecting.  The massless fundamental matters including the chiral fermions are all from the open strings connecting different D6 branes at the interesting points before the spontaneous symmetry breaking. After the symmetry breaking, some of these open strings become massive. So we can model the relevant features of the vacuum picture from the perspective of this brane construction as two D3 put on top of each other before the symmetry breaking, having a U(2) gauge symmetry.  When the two D3 has a separation, this in spirit corresponds to the spontaneous symmetry breaking.  Note also that we can only apply the electric and magnetic fields on the large (1 + 3)-dimensions in the brane construction. This is consistent with that we apply the electric and magnetic fields on one of the D3 branes, taking as the visible (1 + 3)-dimensional world.

In this paper, we will demonstrate, using the simplified system of two D3, that though the stringy computations agree, as expected,  with the corresponding QED computations in the weak-field limit,  the former reveals far more information than the latter, such as the existence of extra dimension(s) and a source of dark matter, if the intersecting D6 brane construction of particle physics standard model is indeed relevant to our real world.  If this is true,   the QED Schwinger pair production can provide a means to detect the existence of the extra dimension(s) and at the same time gives rise to a source of dark matter.

This paper is organized as follows. In  the following  section, we will provide the basic setup for the discussions of the following sections and in particular we will discuss the condition to validate the pair production rate computations.  In section 3, we will move to demonstrate explicitly that the the non-perturbative stringy rate computed agrees completely with the corresponding QED rate in the weak-field limit. We will further spell out the remarkable implications of this agreement.  We will conclude this paper in section 4. 
 
 \section{The basic setup}
 
In order to compute the non-perturbative open string pair production rate, we need first to compute the open string one-loop annulus amplitude between the two D3 branes.   We start with considering each D3 carrying a colinear electric and magnetic fields.  In other words, we will consider the following fluxes on each of the D3 first
 
\be\label{flux}
\hat F^{(a)}_{3} = \left(\begin{array}{cccc}
0& -\hat f_{a}&0&0\\
\hat f_{a}&0&0&\\
0&0& 0 & - \hat g_{a}\\
0&0& \hat g_{a} &0\end{array}\right),
\ee
where both $\hat F^{(a)}_{3}$ are dimensionless with $a = 1, 2$, defined via $\hat F^{(a)} = 2 \pi \alpha' F^{(a)}$ with $F^{(a)}$ the usual field strength tensor. 

The open string one-loop annulus amplitude can then be computed following  \cite{Lu:2017tnm, Lu:2018suj, Jia:2019hbr}\footnote{We actually use the boundary state representation of a D-brane to first compute the closed string tree cylinder amplitude and then express the amplitude in terms of modular forms. After that, we use the Jacobi transformation of various $\theta$-functions to convert this amplitude to the open string one-loop annulus one.}  as
\be \label{annulus-amplit}
\Gamma_{3, 3} =   \frac{4  V_4 |\hat f_1 - \hat f_2| |\hat g_1 - \hat g_2|}{(8 \pi^2 \alpha')^2} \int_0^\infty \frac{d t}{t} e^{- \frac{y^2 t}{2\pi \alpha'}} \frac{(\cosh \pi \nu'_0 t - \cos\pi \nu_0 t)^2}{\sin\pi \nu_0 t \,\sinh\pi \nu'_0 t} Z (t; \nu_{0}, \nu'_{0}),
\ee
where $y$ is the brane separation and 
\be
Z(t;  \nu_{0}, \nu'_{0}) = \prod_{n = 1}^{\infty} \frac{\prod_{j =1}^2 [1 - 2 \,e^{(-)^j \pi \nu'_0 t} |z|^{2n} \cos\pi \nu_0 t + e^{(-)^j 2\pi \nu'_0 t} |z|^{4n}]^2}{(1 - |z|^{2n})^4 (1 - 2\, |z|^{2n} \cos2\pi\nu_0 t + |z|^{4n}) \prod_{j = 1}^2(1 - e^{(-)^{(j - 1)}2\pi\nu'_0 t} |z|^{2n})},
\ee
with $|z| = e^{- \pi t}$.  In the above, the parameters $\nu_{0} \in [0, \infty)$ and $\nu'_{0} \in [0, 1)$ are  determined via
\be\label{egparameter}
\tanh\pi\nu_{0} = \frac{|\hat f_{1} - \hat f_{2}|}{1 - \hat f_{1} \hat f_{2}}, \quad \tan \pi \nu'_{0} = \frac{|\hat g_{1} - \hat g_{2}|}{1 + \hat g_{1}\hat g_{2}}.
\ee
Note that $\nu_{0} \to \infty$ when either $|{\hat f}_{a}| \to 1$, the critical value of electric flux, which can be determined from the Born-Infeld factor used in the boundary state representation of D-brane carrying worldvolume flux, for example, in \cite{DiVecchia:1999uf}. The value of this critical electric field depends on the actual situation. For example, in the holographic case, this was determined from the corresponding Born-Infeld action in \cite{Semenoff:2011ng}.

 For large $t$, the integrand of the above amplitude behaves like $e^{- \frac{t}{2\pi\alpha'} (y^{2} - 2 \pi^{2} \alpha' \nu'_{0})}/t$ which vanishes when $y \ge \pi \sqrt{2 \nu_{0}' \alpha'}$ but blows up when $y <   \pi \sqrt{2 \nu_{0}' \alpha'}$ if $\nu'_{0} \neq 0$, indicating a tachyonic instability at this separation. 
Note also that this amplitude vanishes as expected when $\hat f_{1} = \hat f_{2}$ and $\hat g_{1} = \hat g_{2}$ or in the absence of the electric and magentic fluxes since the underlying system is 1/2 BPS. In general, we have $\Gamma_{3. 3} \neq 0$ and the system breaks all supersymmetries due to the presence of the fluxes. 

It is clear that the above amplitude has an infinite many simple poles occurring when $\sin\pi \nu_{0} t = 0,\, \nu'_{0} \neq 0$ or $\sin\pi \nu_{0} t = 0, \, \cos\pi\nu_{0} t \neq 1, \, \nu'_{0} = 0$ along the positive t-axis. For the former case, these poles can be determined at
\be\label{f-case}
\pi \nu_{0} t_{k} = k \pi \to t_{k} = \frac{k}{\nu_{0}}, \quad k = 1, 2, \cdots,
\ee
while for the latter, the poles occur at
\be
\pi \nu_{0} t_{k} = (2 k - 1) \pi \to t_{k} = \frac{2 k - 1}{\nu_{0}}, \quad k = 1, 2, \cdots.
\ee
The system will relax itself or decay to a stable 1/2 BPS one by producing open string pairs at these poles to lower its excess energy due to the presence of electric fluxes or via tachyonic condensation in the presence of pure magnetic fluxes under the attractive interaction. For the pair production case,  the presence of an infinite number of simple poles implies that the amplitude has an imaginary part and this will give the decay rate of the underlying system.  This decay rate  can be computed as the sum of the residues of the integrand in (\ref{annulus-amplit}) at these poles times $\pi$ per unit worldvolume following  \cite{Bachas:1992bh} as
\bea\label{decayrate}
 {\cal W} &=& - \frac{2 \,{\rm Im} \Gamma}{V_4} \nn
 &=& \frac{8   |{\hat f}_1 - {\hat f}_2||{\hat g}_1 - {\hat g}_2|}{(8\pi^2 \alpha')^2} \sum_{k = 1}^\infty (-)^{k - 1} \frac{\left[\cosh\frac{\pi k \nu'_0}{\nu_0} - (-)^k\right]^2}{k \sinh \frac{\pi k \nu'_0}{\nu_0}} \, e^{- \frac{ k y^2}{2\pi \alpha' \nu_0}} Z_{k} (\nu_{0}, \nu'_{0}),
\eea
where 
\be\label{fn}
Z_{k} (\nu_{0}, \nu'_{0})  = \prod_{n = 1}^{\infty} \frac{\left(1 -  (-)^k \, e^{- \frac{2 n k \pi}{\nu_0} (1 - \frac{\nu'_0}{2 n})}\right)^4 \left(1 - (-)^k  \, e^{- \frac{2 n k \pi}{\nu_0}(1 + \frac{\nu'_0}{2 n})}\right)^4}{\left(1 - e^{- \frac{2 n k \pi}{\nu_0}}\right)^6 \left(1 -   \, e^{- \frac{2 n k \pi}{\nu_0} (1 - \nu'_0/ n)}\right) \left(1 -  \, e^{- \frac{2 n k \pi}{\nu_0}(1 + \nu'_0 / n)}\right)}.
\ee
The open string pair production rate is however different\footnote{After the pairs produced, certain fraction of the pairs will annihilate afterwards and only the average pairs produced per unit time and per unit volume are more relevant for detection and give to the  so-called pair production rate \cite{nikishov}.} and happens to be given as the $k = 1$ term of the above decay rate, following \cite{nikishov} (see the recent discussion in \cite{Lu:2023jxe}). It is
  \be \label{pprate} {\cal W}^{({\rm String})} = 
\frac{8 \,|\hat f_1 - \hat f_2||\hat g_1 - \hat g_2|}{(8\pi^2\alpha')^{2} }  e^{- \frac{y^2}{2\pi \nu_0 \alpha'}}  \frac{\left[\cosh
\frac{\pi \nu'_0}{\nu_0} + 1 \right]^2}{ \sinh
\frac{\pi \nu'_0}{\nu_0}} Z_{1} (\nu_{0}, \nu'_{0}),
\ee
where
\bea\label{z1}
Z_{1} (\nu_{0}, \nu'_{0}) &=&  \prod_{n = 1}^\infty
\frac{\left[1 + 2  e^{- \frac{2 n  \pi}{\nu_0}} \cosh
\frac{\pi \nu'_0}{\nu_0} + e^{- \frac{4 n 
\pi}{\nu_0}}\right]^4}{\left[1 - e^{- \frac{2 n 
\pi}{\nu_0}}\right]^6 \left[1 - e^{- \frac{2  \pi}{\nu_0}(n -
\nu'_0)}\right]\left[1 - e^{- \frac{2  \pi}{\nu_0}(n +
\nu'_0)}\right]}\nn
&=& 1 +  4 \left[1 + \cosh \frac{\pi \nu'_{0}}{\nu_{0}}\right]^{2} \, e^{- \frac{2 \pi}{\nu_{0}}} + \cdots.
\eea
It is clear that when $\hat f_{1} = \hat f_{2}$ and $\hat g_{1} = \hat g_{2}$ (now $\nu_{0} = 0, \nu'_{0} = 0$ from (\ref{egparameter})), $Z_{1} (0, 0) = 1$ and the rate ${\cal W}^{\rm string}$ vanishes as expected.  This counts also the case of an isolated D3 mentioned in the introduction since the two ends of the virtual open string experience the same electric and magnetic fields.

In practice $\nu_{0} \ll 1$,   we only need to keep the leading term contribution of $Z_{1} (\nu_{0}, \nu'_{0})$ to the rate from its expansion given in (\ref{z1}).  The resulting rate is then
 \be \label{pprate-new} {\cal W}^{({\rm String})} = 
\frac{8 \,|\hat f_1 - \hat f_2||\hat g_1 - \hat g_2|}{(8\pi^2\alpha')^{2} }  e^{- \frac{y^2}{2\pi \nu_0 \alpha'}}  \frac{\left[\cosh
\frac{\pi \nu'_0}{\nu_0} + 1 \right]^2}{ \sinh
\frac{\pi \nu'_0}{\nu_0}}.
\ee
In other words, we need to consider the lowest modes of the open string along with the corresponding ones from the anti open string, which consist of eight massive bosonic degrees of freedom (DOF) $8_{\rm B}$ and eight massive fermionic DOF $8_{\rm F}$ from the 10-dimensional view. For the open string, each of the modes is charged, say, with a positive unit charge with respective to the U (1) of the visible D3 and a negative unit charge with respective to that of the hidden D3. For the anti open string, we have the opposite for the charge carried by each mode with respective to the visible D3 and to the hidden D3. Note that all these modes have the same mass given by $m = T_{f} y = y/(2\pi \alpha')$ due to that in the absence of the brane fluxes the system is 1/2 BPS.  So we have in total 16 charged/anti-charged pairs to give rise to the pair production rate (\ref{pprate-new}). These 16 pairs consist of 5  massive scalar pairs, one massive vector pair and 4  massive spinor pairs from either the visible or the hidden D3 worldvolume view.

As stressed in the introduction, the system considered in this paper is a simplified version of the realistic QED constructed from the intersecting D6 branes. Our intention here is to show the key features of such a D-brane construction of particle physics standard model,  which cannot be learned from the usual (1 + 3)-dimensional QED. We will show these features in the following section.  Before closing this section, we need to keep in mind that the string coupling needs to be small to validate the computations for the open string one-loop annulus amplitude and also the rates computed. This is to say that the gauge coupling $g^{2}_{\rm YM} = 2 \pi g_{s}$ is also small which is consistent with the QED charge $e = \sqrt{4 \pi \alpha} = g_{\rm YM}$ with  $\alpha = 1/137$ the fine structure constant. So for our purpose here, we will not have a strong coupling issue as in the holographic case discussed in \cite{Semenoff:2011ng}.

\section{The stringy rate vs the QED rate}

We have no control on the electric and magnetic fields on the hidden D3 and also for simplicity, we set $\hat f_{2} = 0$ and $\hat g_{2} = 0$.  We now express $\hat f_{1} = 2 \pi \alpha' e E$ and $\hat g_{1} = 2 \pi \alpha' e B $ in terms of the usual electric field $E$ and magnetic field $B$ on the visible D3. So we have from (\ref{pprate-new})
 \be\label{eg3pprate-new1}
{\cal W}^{({\rm String})} = \frac{ 2 (e E) (e B)}{(2 \pi)^2 } \frac{\left(\cosh\frac{\pi B}{ E} +1\right)^2}{\sinh \frac{\pi B}{E}} \, e^{- \frac{  \pi m^{2} (y)}{e E }},
\ee
where we have used (\ref{egparameter}) to  express $\nu_{0}$ and $\nu'_{0}$ in terms of $E$ and $B$ noticing $\hat f_{1} \ll 1$ and $\hat g_{1} \ll 1$ in practice and the mass scale $m$ 
\be\label{mass}
m (y) =  T_{F} y = \frac{y}{2\pi \alpha'},
\ee
as given earlier.  The rate formula (\ref{eg3pprate-new1}) is our starting point in what follows.

 Note that this rate is a non-perturbative one which can be seen from the appearance of the coupling $e = g_{\rm YM} = \sqrt{2\pi g_{s}}$ in the exponential factor in the formula (\ref{eg3pprate-new1}).  In other words, any perturbative expansion of this formula around $e = 0$ gives a null result.

The QED non-perturbative rates for a scalar charged pair and a spinor charged pair, respectively, with the same applied electric and magnetic fields were given long time ago by Nikishov \cite{nikishov} (see also \cite{Kim:2003qp}) as 
 \be\label{scalar/spinor}
 {\cal W}^{\rm QED}_{\rm scalar} = \frac{(e E) (e B)}{2 (2\pi)^{2}} {\rm csch} \left(\frac{\pi B}{E}\right) \, e^{- \frac{\pi m^{2}}{e E}},\quad
{\cal W}^{\rm QED}_{\rm spinor} = \frac{(e E) (e B)}{(2\pi)^{2}} \coth\left(\frac{\pi B}{E}\right) \, e^{- \frac{\pi m^{2}}{e E}},
\ee 
while for a vector pair, it is \cite{Kruglov:2001cx}
 \be\label{w-bosonr}
{\cal W}^{\rm QED}_{\rm vector} = \frac{(e E) (e B)}{ 2 (2 \pi)^{2}} \frac{ 2 \cosh \frac{2 \pi B}{E} + 1}{ \sinh \frac{\pi B}{E}} \, e^{- \frac{\pi m^{2}}{e E}}.
\ee

Given that the stringy rate (\ref{eg3pprate-new1}) is due to the contributions from five scalar pairs, four spinor pairs and one vector pair from the viewpoint of observer on the visible D3 brane,  one expects that this rate should be identical to the following QED one  
\be\label{QED-rate}
{\cal W}^{\rm QED} = 5\, {\cal W}^{\rm QED}_{\rm scalar} + 4\, {\cal W}^{\rm QED}_{\rm spinor} + {\cal W}^{\rm QED}_{\rm vector},  
\ee
if we set the mass $m$ here to be the same as the mass $m (y)$ as given in (\ref{eg3pprate-new1}).  One happily finds that this is indeed true, i.e.,
\be
{\cal W}^{({\rm String})}  = {\cal W}^{\rm QED}.
\ee
This identity, though expected, is remarkable in the following sense. The stringy rate ${\cal W}^{\rm string}$ is computed non-perturbatively using the stringy description of D-branes and in particular its non-vanishing value needs the presence of a second (hidden) D3 (for an isolated D3, this rate is zero). Further the modes contributing to this rate come from the open string/anti open string pair connecting the two D3 branes under consideration, therefore along the directions transverse to both of the two D3 branes, i.e., along the extra dimension(s) from the worldvolume viewpoint. In other words, a non-vanishing rate implies the existence of extra dimension(s) and also the existence of the hidden brane nearby our visible D3 brane.  On the other hand, the QED rate ${\cal W}^{\rm QED}$ needs only our (1 + 3)-dimensional world and the corresponding pairs contributing to this rate.  It says nothing about the hidden D3 and the existence of extra dimension(s).  

In other words, the low energy effective quantum field theory can give correct computations if the low energy modes are correctly identified, but cannot reveal more than the visible (1 + 3)-dimensional world physics. So if the intersecting D-brane construction of particle physics standard model is indeed relevant to our real world, a detection of the QED Schwinger pair production provides a means to detect the existence of extra dimension(s) and also a means to detect the existence of dark matter as a form of hidden branes (not in the form of particles) nearby our visible branes in the extra dimensions.

\section{Conclusion and discussion}
In this paper, the  null detection of the electron-positron pair production in QED so far and the upcoming experiments for such a detection motivate us to check if the QED vacuum picture is  complete or not. There is a certain rationality for posing such a question since from the modern view,  QED, though very successful,  is just a low energy effective description of the underlying fundamental theory and so is its underlying vacuum. 

 For this, we use a simplified version of the intersecting D brane construction of particle physics standard model by considering a system of two D3 placed parallel at a separation. One of the two D3 is taken as the visible (1 + 3)-dimensional world, mimicking our own (1 + 3)-dimensional one, while the other D3, dark to observers in the visible one, is a hidden one.  This simplified system captures the essential features of the QED vacuum from the brane construction.  The presence of the hidden D3 is a must for having a non-vanishing pair production (as discussed in Introduction an isolated D brane cannot have a non-vanishing pair production if the applied electric field is below the critical one).  The open string/anti open string pair produced, though  only the pairs coming from their lowest modes contribute in the weak-field limit,   is from the one connecting these two D3. In other words, if there is any pair production, it must come from this kind of open string /anti open string pairs, signaling the existence of extra dimension(s). While for QED computations, the same rate can be obtained for the same low energy pairs resulting from the quantum fluctuations of the corresponding QED vacuum but here we  need only the (1 + 3)-dimensional world, nothing else.

So there is a clear difference regarding the underlying vacuum picture between the stringy description and the QED one. This provides us a means to detect the existence of extra dimension(s) and gives rise to a source of dark matter (the hidden branes), not in the form of particles, if the intersecting D brane construction of particle standard model is  indeed relevant\footnote{Given that the matter needs to be in the fundamental representation of standard model groups, this is not so surprised since in the D brane construction, the fundamental representation comes from that the two ends of an open string attach on different D branes.}.

So a future detection of the QED Schwinger pair production  can teach us potential lessons about the existence of extra dimension(s) and a source of dark matter.  If on the other hand, we still have a null detection even for a large enough laboratory electric field $E \ge E_{T}$, this must imply that the QED vacuum picture has a serious problem. Though the isolated D brane appears to give an explanation of this qualitatively, one has to explain where the fundamental matter comes from in the brane construction.


\section*{Acknowledgments}
The author would like to thank Xu-Guang Huang,  Tianjun Li, Li Lin Yang, Mirjam Cvetic, Kai Chang, Li You, Hong Lu, Zhao-Long Wang for useful discussions at various stages of this manuscript.  We acknowledge the support by grants from the NNSF of China with Grant No: 12275264 and 12247103.


\begin{thebibliography}{99}

\bibitem{Schwinger:1951nm}
  J.~S.~Schwinger,
  ``On gauge invariance and vacuum polarization,''
  Phys.\ Rev.\  {\bf 82}, 664 (1951).
   
 \bibitem{PS1945}
 I.~Pomeranchuk and J. Smorodinsky,   J. Phys. USSR 9, 97 (1945).
 
 \bibitem{GZ1969}
 S.~ S.~ Gershtein and Y.~ B.~ Zeldovich, Zh. Eksp. Teor. Fiz. 57,
654 (1969) [Sov. Phys. JETP 30, 358 (1970)]; Lett. Nuovo
Cimento 1, 835 (1969).

\bibitem{PG1969}
 W. Pieper and W. Greiner, Z. Phys. 218, 327 (1969).
 
\bibitem{ZP1972}
 Y.~ B.~ Zeldovich and V. S. Popov, Sov. Phys. Usp. 14, 673 (1972).
 
\bibitem{GMR1985}
 W.~ Greiner, B.~ Müller, and J.~ Rafelski, Quantum Electrodynamics of Strong Fields (Springer-Verlag, Berlin, 1985).
 
 \bibitem{MS1994}
  U.~ Müller-Nehler and G.~ Soff, Phys. Rep. 246, 101 (1994).
 
\bibitem{GMV2017}
S.~ I.~ Godunov, B.~ Machet, and M.~ I.~ Vysotsky, Eur. Phys. J. C 77, 782 (2017).

\bibitem{ahmad1999}
I.~ Ahmad et al., Phys. Rev. C 60, 064601 (1999). 

\bibitem{exp1-2009}
 A.~ Gumberidze, Th.~ Stöhlker, H.~ F.~ Beyer, F.~ Bosch, A.~
Bräuning-Demian, S.~ Hagmann, C.~ Kozhuharov, Th.~ Kühl,
R.~ Mann, P.~ Indelicato, W.~ Quint, R.~ Schuch, and A.~
Warczak, Nucl. Instrum. Methods Phys. Res., Sect. B 267, 248 (2009).

\bibitem{exp2-2015}
G.~ M.~ Ter-Akopian, W.~ Greiner, I.~ N.~ Meshkov, Y.~ T.~
Oganessian, J.~ Reinhardt, and G.~ V.~ Trubnikov, Int. J. Mod. Phys. E 24, 1550016 (2015).

\bibitem{exp-3-2017}
X.~ Ma, W.~ Q.~ Wen, S.~ F.~ Zhang, D.~ Y. ~Yu, R.~ Cheng, J.~ Yang,
Z.~ K.~ Huang, H.~ B.~ Wang, X.~ L.~ Zhu, X.~ Cai, Y.~ T.~ Zhao,
L.~ J.~ Mao, J.~ C.~ Yang, X.~ H.~ Zhou, H.~ S.~ Xu, Y.~ J.~ Yuan, J.~ W.~
Xia, H.~ W.~ Zhao, G.~ Q.~ Xiao, and W.~ L. ~Zhan, Nucl. Instrum. Methods Phys. Res., Sect. B 408, 169 (2017).

\bibitem{Maltsev:2019ytv}
I.~A.~Maltsev, V.~M.~Shabaev, R.~V.~Popov, Y.~S.~Kozhedub, G.~Plunien, X.~Ma, T.~St\"ohlker and D.~A.~Tumakov,
``How to observe the vacuum decay in low-energy heavy-ion collisions,''
Phys. Rev. Lett. \textbf{123}, no.11, 113401 (2019)
doi:10.1103/PhysRevLett.123.113401
[arXiv:1903.08546 [physics.atom-ph]].

\bibitem{Popov:2020xmd}
R.~V.~Popov, V.~M.~Shabaev, D.~A.~Telnov, I.~I.~Tupitsyn, I.~A.~Maltsev, Y.~S.~Kozhedub, A.~I.~Bondarev, N.~V.~Kozin, X.~Ma and G.~Plunien, \textit{et al.}
``How to access QED at a supercritical Coulomb field,''
Phys. Rev. D \textbf{102}, no.7, 076005 (2020)
doi:10.1103/PhysRevD.102.076005
[arXiv:2008.05005 [hep-ph]].
 
  \bibitem{smf}
  S. Hahn, Kwanglok Kim, Kwangmin, Kim, X. Hu, T. Painter, I. Dixon,  S.  Kim,  K. R. Bhattarai, S. Noguchi, J. Jaroszynski and D. C. Larbalestier,
  ``45.5-tesla direct-current magnetic field generated with a high-temperature superconducting magnet,'' Nature {\bf 570}, 496-499 (2019).
  
  
\bibitem{Lu:2018nsc}
  J.~X.~Lu,
  ``A possible signature of extra-dimensions: The enhanced open string pair production,''
  Phys.\ Lett.\ B {\bf 788}, 480 (2019)
  [arXiv:1808.04950 [hep-th]].
  
  
\bibitem{Jia:2018mlr}
  Q.~Jia and J.~X.~Lu,
  ``Remark on the open string pair production enhancement,''
  Phys.\ Lett.\ B {\bf 789}, 568 (2019)
  [arXiv:1809.03806 [hep-th]].

  
\bibitem{Lu:2020hml}
J.~X.~Lu and N.~Zhang,
`More on the open string pair production,''
Nucl. Phys. B \textbf{977}, 115721 (2022)
doi:10.1016/j.nuclphysb.2022.115721
[arXiv:2002.09940 [hep-th]].

\bibitem{Lu:2023jxe}
J.~X.~Lu,
``Understanding the open string pair production of the Dp/D0 system,''
JHEP \textbf{11}, 019 (2023)
doi:10.1007/JHEP11(2023)019
[arXiv:2307.06594 [hep-th]].

\bibitem{Lu:2023sag}
J.~X.~Lu,
``The open string pair production, its enhancement and the physics behind,''
Phys. Lett. B \textbf{848}, 138397 (2024)
doi:10.1016/j.physletb.2023.138397
[arXiv:2310.07960 [hep-th]].

\bibitem{Bachas:1992bh}
  C.~Bachas and M.~Porrati,
  ``Pair creation of open strings in an electric field,''
  Phys.\ Lett.\  B {\bf 296}, 77 (1992)
  [arXiv:hep-th/9209032].

\bibitem{Porrati:1993qd}
  M.~Porrati,
  ``Open strings in constant electric and magnetic fields,''
  arXiv:hep-th/9309114.
  
\bibitem{Gorsky:2001up}
A.~S.~Gorsky, K.~A.~Saraikin and K.~G.~Selivanov,
``Schwinger type processes via branes and their gravity duals,''
Nucl. Phys. B \textbf{628}, 270-294 (2002)
doi:10.1016/S0550-3213(02)00095-0
[arXiv:hep-th/0110178 [hep-th]].

\bibitem{Semenoff:2011ng}
G.~W.~Semenoff and K.~Zarembo,
``Holographic Schwinger Effect,''
Phys. Rev. Lett. \textbf{107}, 171601 (2011)
doi:10.1103/PhysRevLett.107.171601
[arXiv:1109.2920 [hep-th]].

  


\bibitem{Lu:2017tnm}
  J.~X.~Lu,
  ``Magnetically-enhanced open string pair production,''
  JHEP {\bf 1712}, 076 (2017)
  doi:10.1007/JHEP12(2017)076
  [arXiv:1710.02660 [hep-th]].

\bibitem{Lu:2018suj}
  J.~X.~Lu,
  ``Some aspects of interaction amplitudes of D branes carrying worldvolume fluxes,''
  Nucl.\ Phys.\ B {\bf 934}, 39 (2018)
  [arXiv:1801.03411 [hep-th]].
  
    \bibitem{Lu:2019ynq}
  J.~X.~Lu,
  ``A note on the open string pair production of the D3/D1 system,''
  JHEP {\bf 1910}, 238 (2019)
  doi:10.1007/JHEP10(2019)238
  [arXiv:1907.12637 [hep-th]].
  
  
\bibitem{Ibanez:2001nd}
L.~E.~Ibanez, F.~Marchesano and R.~Rabadan,
``Getting just the standard model at intersecting branes,''
JHEP \textbf{11}, 002 (2001)
doi:10.1088/1126-6708/2001/11/002
[arXiv:hep-th/0105155 [hep-th]].

\bibitem{Cvetic:2001nr}
M.~Cvetic, G.~Shiu and A.~M.~Uranga,
``Chiral four-dimensional N=1 supersymmetric type 2A orientifolds from intersecting D6 branes,''
Nucl. Phys. B \textbf{615}, 3-32 (2001)
doi:10.1016/S0550-3213(01)00427-8
[arXiv:hep-th/0107166 [hep-th]].

\bibitem{Blumenhagen:2005mu}
R.~Blumenhagen, M.~Cvetic, P.~Langacker and G.~Shiu,
``Toward realistic intersecting D-brane models,''
Ann. Rev. Nucl. Part. Sci. \textbf{55}, 71-139 (2005)
doi:10.1146/annurev.nucl.55.090704.151541
[arXiv:hep-th/0502005 [hep-th]].

\bibitem{DiVecchia:1999uf}
P.~Di Vecchia, M.~Frau, A.~Lerda and A.~Liccardo,
``(F,D(p)) bound states from the boundary state,''
Nucl. Phys. B \textbf{565}, 397-426 (2000)
doi:10.1016/S0550-3213(99)00632-X
[arXiv:hep-th/9906214 [hep-th]].

\bibitem{Jia:2019hbr}
Q.~Jia, J.~X.~Lu, Z.~Wu and X.~Zhu,
``On D-brane interaction \& its related properties,''
Nucl. Phys. B \textbf{953}, 114947 (2020)
doi:10.1016/j.nuclphysb.2020.114947
[arXiv:1904.12480 [hep-th]].


  
  \bibitem{nikishov}
  A. I. Nikishov, Sov. Phys. JETP 30, 660 (1970); Nucl. Phys. B21, 346 (1970).
  
\bibitem{Kim:2003qp}
S.~P.~Kim and D.~N.~Page,
``Schwinger pair production in electric and magnetic fields,''
Phys. Rev. D \textbf{73}, 065020 (2006)
doi:10.1103/PhysRevD.73.065020
[arXiv:hep-th/0301132 [hep-th]].
  
\bibitem{Kruglov:2001cx} 
  S.~I.~Kruglov,
  ``Pair production and vacuum polarization of vector particles with electric dipole moments and anomalous magnetic moments,''
  Eur.\ Phys.\ J.\ C {\bf 22}, 89 (2001)
  [hep-ph/0110100].  




   \end{thebibliography}
\end{document}